\documentclass[prl,twocolumn]{revtex4}
\usepackage{epsfig,amssymb,amsmath,graphicx,enumerate,color}
\definecolor{MyDarkBlue}{rgb}{0.1,0,0.55} 
\pdfoutput = 1

\begin{document}

\title{Optical codeword demodulation with error rates below standard quantum limit using a conditional nulling receiver}

\author{Jian Chen, Jonathan L. Habif, Zachary Dutton, Richard Lazarus, Saikat Guha}
\affiliation{Disruptive Information Processing Technologies group, Raytheon BBN Technologies, Cambridge, MA 02138, USA}

\begin{abstract}
The quantum states of two laser pulses---coherent states---are never mutually orthogonal, making perfect discrimination impossible. Even so, coherent states can achieve the ultimate quantum limit for capacity of a classical channel, the Holevo capacity. Attaining this requires the receiver to make joint-detection measurements on long codeword blocks, optical implementations of which remain unknown. We report the first experimental demonstration of a joint-detection receiver, demodulating quaternary pulse-position-modulation (PPM) codewords at a word error rate of up to 40\% ($2.2$~dB) below that attained with direct-detection, the largest error-rate improvement over the standard quantum limit reported to date.   This is accomplished with a conditional nulling receiver, which uses optimized-amplitude coherent pulse nulling, single photon detection and quantum feedforward.  We further show how this translates into coding complexity improvements for practical PPM systems, such as in deep-space communication. We anticipate our experiment to motivate future work towards building Holevo-capacity-achieving joint-detection receivers.
\end{abstract}

\maketitle

One of the most important insights of quantum physics in the modern theory of optical communication is the realization that it is {\em impossible} to perfectly discriminate states of light that are not mutually orthogonal. While orthogonal quantum states of light $\left\{|\psi_k\rangle\right\}$, $k=1, 2, \ldots$, i.e., whose inner products satisfy $\langle{\psi_k}|{\psi_j}\rangle = \delta_{kj}$, {\em can} in principle be discriminated with zero probability of error, such states of light are hard to create~\footnote{One example of a class of orthogonal states are the so-called Fock states of light $\left\{|0\rangle, |1\rangle, \ldots\right\}$ (quantum light pulses with a fixed number of photons), which can in principle be discriminated with a perfect {\em photon number resolving} (PNR) detector. Whereas both Fock states and PNR detectors are useful resources for quantum information processing, they are both extremely hard to realize in practice.}. An ordinary laser pulse is in a {\em coherent state}, $|\alpha\rangle$ (where $\alpha$ is a complex number denoting the mean field value). However, no two coherent states---even those in orthogonal space-time field modes---are ever in mutually orthogonal quantum states, i.e., $\langle\alpha{|}\beta\rangle = \exp\left[\alpha^*\beta - \frac12{\left(|\alpha|^2 + |\beta|^2\right)}\right] \ne 0$, precluding perfect discrimination. In spite of this apparent impairment, coherent states are sufficient to attain the ultimate (Holevo) capacity of optical communication, even on a lossy channel~\cite{Gio04}, a channel on which mutually-orthogonal quantum states such as Fock states actually fail to achieve the Holevo capacity. Recently, we discovered the first Holevo-capacity achieving codes~\cite{Wil11}, and some advances on the optimal receiver measurements have been reported in recent literature~\cite{Llo10,Guh11a,Wil11}. However, designing and building such optical receivers remain elusive.

Helstrom gave a recipe to compute the minimum achievable average probability of error to discriminate quantum states from a given ensemble~\cite{Hel76}, known as the Helstrom limit. He also found necessary and sufficient conditions on the operators describing the optimal measurement that achieves that minimum. Unfortunately, very little is known in general about how to build receivers achieving this limit using laboratory optics. The binary-hypothesis Dolinar receiver can attain it in the case of discriminating two coherent states~\cite{Dol76}. This was demonstrated only recently, three decades after its invention~\cite{Coo07}, owing to the difficulty of realizing the ultrafast electro-optic feedback and high-bandwidth single-photon detection it requires. In addition, several sub-optimal receiver structures have been proposed which bridge some of the gap between the {\em standard quantum limit} (SQL)---the limit of conventional optical detection strategies for the discrimination tasks---and the Helstrom limit. Examples include Kennedy's~\cite{Ken72} and Takeoka et al.'s~\cite{Tak08, Wit08} receivers to discriminate a binary coherent state constellation and Bondurant's receiver~\cite{Bon93} for the quaternary phase-shift-keying constellation.  In \cite{Tsu11} discrimination was performed better than the SQL obtainable with perfect quantum efficiency detectors.

While all these receivers surpass the SQL for optical state discrimination, they share the feature of acting only on single symbols of the modulation alphabet at a time. However, it is known that joint quantum measurements across multiple symbols are required to achieve the Holevo capacity~\cite{Sas98,Fuc97,Guh11}. The conditional pulse nulling (CPN) receiver to discriminate pulse-position-modulation (PPM) codewords at a word error rate below the SQL (direct detection in this case), was the first structured optical receiver proposal to employ a joint detection strategy~\cite{Dol83}, the first implementation of which is the key accomplishment reported in this article. There have since been several discussions of joint measurements acting on codeword blocks of symbols that can attain fundamentally higher capacity than the best single symbol detector, dubbed ``super-additive capacity"~\cite{Sas98, Fuc97}. We recently proposed an improved version of the CPN receiver~\cite{Guh10}, some of the first explicit optical realizations of joint-detection receivers~\cite{Guh11,Guh11a}, and the first explicit Holevo capacity achieving codes~\cite{Wil11}.

Traditional PPM demodulation uses direct detection (DD) of each symbol in the codeword. We denote the quantum state for a single-mode coherent state pulse with a mean of $N_p$ photons by the quantum ket vector $|\alpha\rangle$ with $\alpha = \sqrt{N_p}e^{j\phi}$. Ideal direct detection (photon counting) of this pulse generates clicks in a Poisson point process.  Using a single photon detector (SPD) results in a click occuring or not with probability $P_1 = 1-e^{- \eta N_p}$ and $P_0 = e^{- \eta N_p}$ respectively, where $\eta$ denotes the quantum efficiency of the detector. For discriminating $M$-ary PPM codewords ($M$ codewords, each being a sequence of $M$ coherent state pulses, of the form $|0\rangle \ldots |\alpha\rangle \ldots |0\rangle$, $|\alpha|^2 = N_p$, of $M$ ``pulse positions"), this translates to an average error probability $P_e^{(DD)}=((M-1)/M) e^{- \eta N_p}$.

\begin{figure}
\centering
\includegraphics[width=\columnwidth]{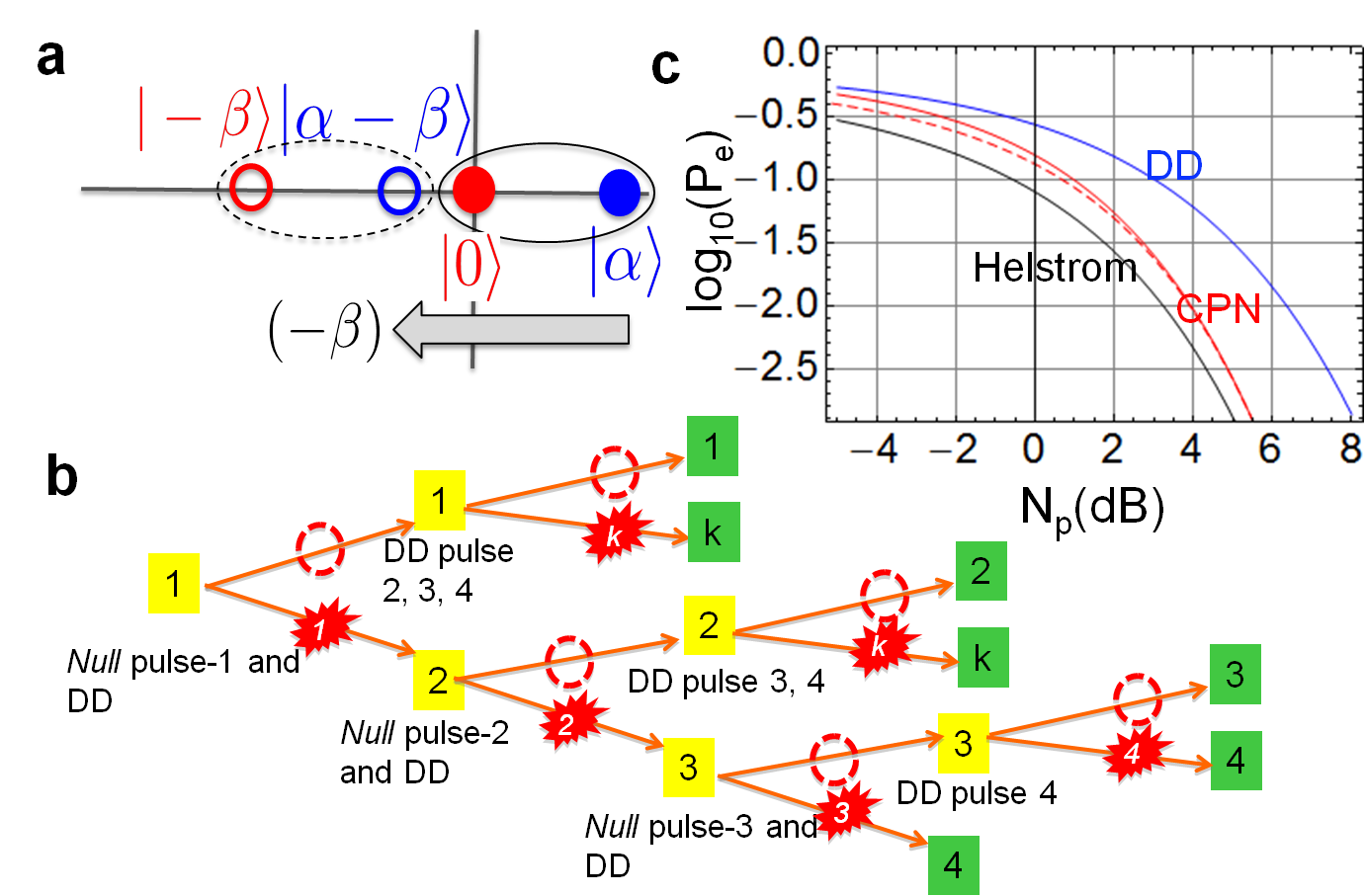}
\caption{ {\bf(a)} An on-off-keying coherent state alphabet $\left\{|0\rangle, |\alpha\rangle\right\}$ is shifted to $\left\{|-\beta\rangle, |\alpha-\beta\rangle\right\}$, by mixing the input coherent state on a beam-splitter with near-unity transmissivity ($\kappa = 0.99$) with a local coherent state $|\beta/\sqrt{1-\kappa}\rangle$. The shifted state is direct-detected with a single photon detector (SPD), resulting in an average probability of error $P_{e,{\rm GK}} = \min_{\beta}\frac12{\left[e^{-(\alpha - \beta)^2} + (1-e^{-\beta^2})\right]}$. In the special case of ``perfect nulling" ($\beta=\alpha$), the output constellation is $\left\{|-\alpha\rangle, |0\rangle\right\}$, which has an effect of swapping the on and off probabilities. This (Generalized Kennedy) receiver is a building block for the conditional pulse nulling (CPN) receiver to demodulate PPM codewords, in which the click records obtained in previous pulse slots determine the nulling amplitude $\beta$ on the current slot. {\bf(b)} The binary decision circuit for the CPN receiver, showed here for the $4$-ary PPM case. The numbers in yellow represent the receiver's current hypotheses, whereas the numbers in green denote the final decision ($k$ denoting the pulse position where a click was detected). The receiver starts out by hypothesizing that code-word $1$ was sent and nulls the first time slot in order to displace the hypothesized pulse to vacuum.  A no-click event (upper-branch) supports the correct hypothesis and the receiver reverts to direct detection thereafter. A click (lower branch) indicates an incorrect hypothesis and the receiver then nulls time slot two. A similar series of binary decisions occurs at each time slot. {\bf(c)} Average probability of error for 4-ary PPM codeword discrimination versus mean photon number per pulse $N_p$, using direct detection (blue), the perfect nulling CPN receiver (red:solid), the optimal nulling CPN receiver (red:dashed), and the Helstrom bound (black). Note that the error exponent of the Helstrom limit ($e^{-2N_p}$) is twice that of the direct-detection receiver ($e^{-N_p}$).}
\label{fig:CPN}
\end{figure}

The CPN receiver improves upon this by combining two strategies: (1) manipulation of the quantum (Poisson) noise statistics via coherent pulse nulling; and (2) updating the nulling strategy in each time slot based on detection events in previous time slots. The first of these is essentially the advantage realized in the single-symbol Generalized Kennedy (GK) receiver~\cite{Tak08,Wit08} for discriminating two coherent states. A ``nulling" pulse $|\beta/\sqrt{1-\kappa}\rangle$ is mixed with the incoming signal pulse $|\alpha\rangle$ on a highly asymmetric beam-splitter of transmissivity $\kappa \approx 1$, resulting in a displacement of the signal  $|\alpha\rangle$ to $|\alpha-\beta\rangle$. The key insight of a nulling based receiver is that the ``on-off" detection probability of a coherent state stemming from the Poisson noise of direct detection is a function of its intensity, and can be adjusted via shifting the entire modulation constellation. Fig.~\ref{fig:CPN}(a) illustrates this for the binary {\it on-off-keyed} (OOK) modulation alphabet, where the Generalized Kennedy receiver chooses a displacement $\beta$ to minimize the average probability of error in discriminating $|0\rangle$ and $|\alpha\rangle$. As an example, for $\alpha = 0.2$, $P_{e,\mathrm{DD}}=0.480$, whereas $P_{e,\mathrm{GK}}=0.415$ (attained for $\beta = -0.61$ or 0.71). This is remarkably close to the Helstrom limit, $P_{e,\mathrm{min}}=0.401$. A similar receiver concept has been proposed for the $4$-ary {\it phase-shift-keyed} (4PSK) alphabet~\cite{Bon93}.

The second key feature of the CPN receiver is the joint detection strategy implemented by utilizing quantum feed-forward on successive pulse slots based on detection results from previous time slots.  In contrast to the Dolinar receiver, which requires instant intra-pulse feedback to perform optimally, the CPN receiver only requires inter-pulse feed-forward at a speed commensurate with the pulse repetition rate.  Using the previous pulse click history allows the receiver to update the posterior probabilities of the hypotheses and choose a nulling strategy that minimizes the final codeword probability of error. Specifically, the receiver operates by nulling the $M$ pulse slots sequentially per the decision tree shown in Fig.~\ref{fig:CPN}(b). It starts by applying an {\em exact} nulling optical pulse ($\beta=\alpha$) to the first pulse slot and detecting the nulled slot by an SPD. If a click is not observed in that slot, the receiver continues to believe that the pulse was in fact present in that first time slot, and proceeds to direct detect the remainder of the pulse slots without nulling. However, if the first slot did generate a click, the first slot is rejected as a hypothesis after which the receiver is faced with an $M-1$-ary version of the original discrimination task. It moves to the second slot, nulls it, detects it, and moves on recursively. This pulse-to-pulse feedforward is what enables the receiver to jointly process the information from all pulse slots collectively in a more efficient way than detecting each slot individually. Fig.~\ref{fig:CPN}(c) plots the average codeword discrimination error rates for 4-ary PPM codewords $\left\{|\alpha\rangle|0\rangle|0\rangle{|0\rangle}, |0\rangle|\alpha\rangle|0\rangle{|0\rangle},|0\rangle|0\rangle|\alpha\rangle|0\rangle,|0\rangle|0\rangle|0\rangle{|\alpha\rangle}\right\}$ using DD and CPN, and compares them to the Helstrom limit. At high $N_p$ the CPN error attains an error rate scaling $\sim e^{-2 N_p}$ matching the Helstrom bound while DD attains $\sim e^{-N_p}$. We recently found two improved versions of Dolinar's exact pulse nulling CPN receiver. One uses optimized nulling amplitudes in a similar manner to the Generalized Kennedy receiver described above. The dashed red curve in Fig.~\ref{fig:CPN}(c) shows the improved performance over that of the exact nulling CPN receiver, particularly at low photon numbers~\cite{Guh10}. We implemented both the exact and optimized nulling receivers, as described below.  The other improvement, not implemented here,  is obtained by applying an optimal amount of squeezing (phase-sensitive amplification) on the pulses to further manipulate the quantum noise statistics upon detection.

\begin{figure}
\centering
\includegraphics[width=\columnwidth]{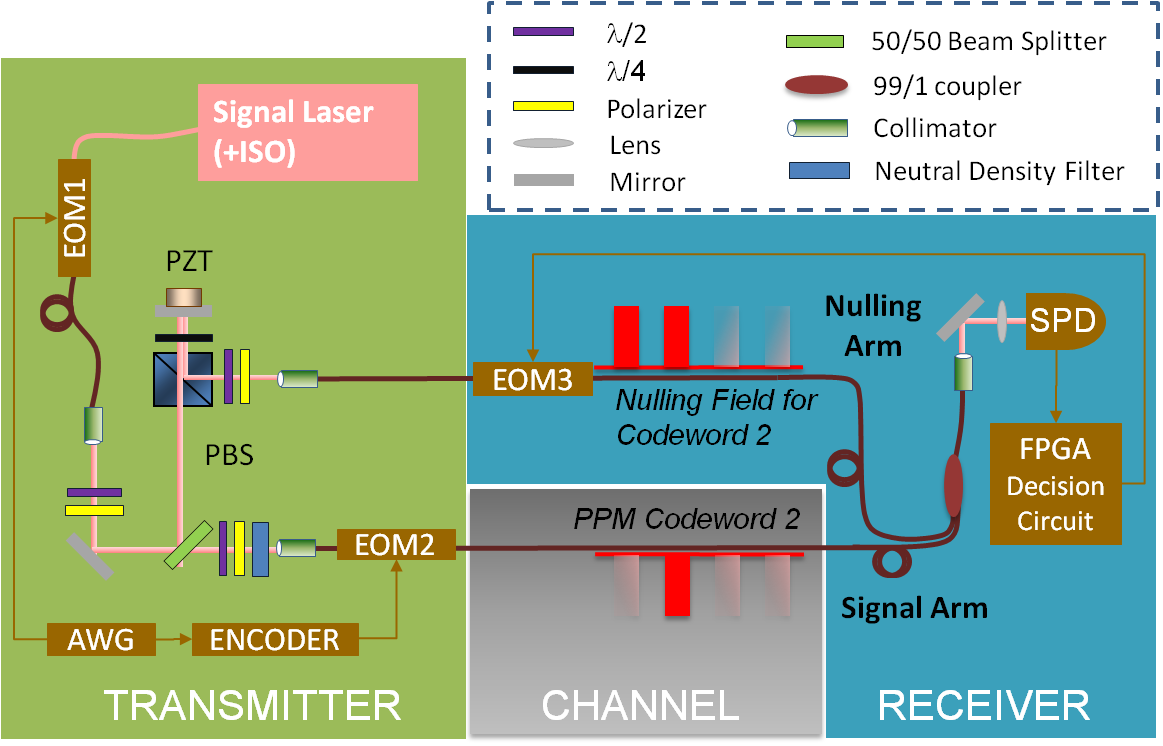}
\caption{The experimental diagram of our CPN receiver implementation. EOM1 carves $50$~ns pulses spaced by 1~$\mu$s and EOM2 (EOM3) implements the pulse pattern for the signal PPM codewords (nulling pulse sequence). The nulling pulse  decision circuit is implemented via an FPGA. The example pulse sequences shown in the two arms represent a signal transmitting codeword $2$ and the corresponding most probable nulling pattern for high $N_p$.}
\label{fig:experiment}
\end{figure}

Our experimental setup is shown in Fig.~\ref{fig:experiment}. Frames of four 50 ns flat-top pulses in each of the four pulse slots are carved from a polarized CW laser beam ($\lambda=$688~nm) by an electro-optic amplitude modulator (EOM1).  Pulse slots within PPM codewords are spaced by $1~\mu$s.  Each is split at a 50/50 beamsplitter and sent into the ``signal arm'' and the ``nulling arm'' where the PPM signal is encoded, and the conditional feedback is prepared, respectively.   Each 4-ary codeword is spaced by $6~\mu$s.  The pulse configurations for the signal (nulling) arm are prepared independently by using EOM2 (EOM3) in each arm as a shutter to select which pulses in the four-pulse train are transmitted. In the signal arm, one of the four pulses is transmitted by EOM2 to encode the desired PPM codeword. In the nulling arm, a field programmable gate array (FPGA) implements the CPN decision tree shown in Fig~\ref{fig:CPN}(d) and controls the transmissivity of EOM3.  A Xilinx Spartan6 FPGA implements the 3-state state machine of the decision tree, monitoring the single photon detector output, and controlling the optical feedback in the nulling arm for the subsequent pulse slot conditioned on the presence or absence of a detection event in the current pulse slot. The resulting optical fields are combined on a 99/1 beam-splitter with high transmissivity in the signal arm. To maintain fully destructive interference between the two fields at the beam-splitter, a piezo-driven mirror is placed in the signal arm to adjust the signal arm path length relative to the nulling arm.  The resulting field is directed towards a silicon avalanche photodiode single photon detector with detection efficiency $\eta = 0.40$, and dark count rate $16$ counts/sec.  A detection event during a pulse slot triggers the decision tree to move to its next branch and either provide nulling to the next pulse slot or not. A representative example of transmitting codeword $2$ and an associated nulling sequence is shown in Fig.~\ref{fig:experiment}.

\begin{figure}
\centering
\includegraphics[width=\columnwidth]{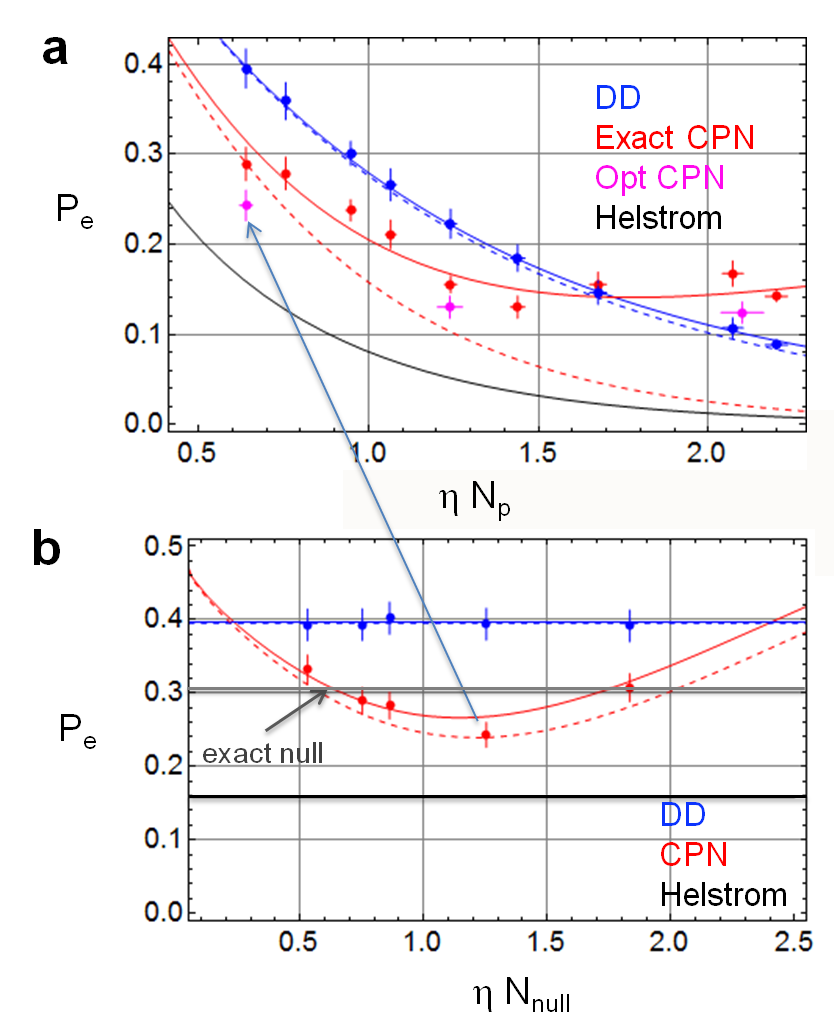}
\caption{{\bf(a)} The probability of error for DD (blue dots) and exact nulling CPN (red dots). Vertical error bars are statistical errors due to finite number of error events out of the 832 frames (some data points were taken more than once resulting in smaller error bars).  Horizontal error bars are calculated from measured intensity fluctuation levels. The dashed curves show the ideal calculated performance of the two receivers, in the absence of dark (``d") and background (``bg") counts, and any mode (``$\Delta_m$") or phase (``$\Delta_\theta$") mis-match. The solid blue curves show a DD model with $P_{\mathrm{d+bg}}=0.0042 N_p+ 0.0129 N_\mathrm{null}$ while the solid red curve is a CPN model with this $P_{\mathrm{d+bg}}$ and with $\Delta_m+\Delta_\theta=0.05$. The magenta dots show the experimental results obtained by optimizing the nulling photon number $N_{\rm null}$ at several different $N_p$ values. {\bf(b)} A nulling photon number ($N_{\rm null}$) sweep holding $N_p=0.64$ constant. The solid curves use the same model as in (a), but the mis-match parameter is chosen to be $\Delta_m+\Delta_\theta=0.03$ to give the best fit to the data.}
\label{fig:error_rates}
\end{figure}

For each of the codewords, we input a set of $832$ frames and decoded them using both DD and the exact nulling CPN receiver. For each PPM frame, the click record over the four slots was used to choose the most likely codeword.  In certain cases, more than one outcome is equally probable (both due to fundamental Poisson noise and errors due to dark clicks and imperfect mode matching of the null and signal), in which case the decoder randomly chooses between words that are equally probable. We then compared the receiver-generated hypothesis with the known input state to calculate the sample-mean probability of error in each case. This was done for a series of different pulse mean photon numbers $N_p$, which was controlled by adjusting a neutral density filter in the signal arm (see Fig.~\ref{fig:experiment}) and measured by processing the click records of the DD data. The results are plotted in Fig.~\ref{fig:error_rates}(a) versus the signal pulse photon number (normalized to include the measured quantum efficiency $\eta=0.40$). We see that the CPN receiver out-performs DD over a large range of $N_p$. The largest fractional improvement of $1.6$~dB over DD is observed at $N_p=1.25$ photons. At higher $N_p$, the CPN error rate becomes worse than that of DD, unlike what is predicted by ideal theory.  The prediction under ideal conditions is shown as the dashed red curve and scales like the Helstrom limit at high $N_p$. 

To  understand the deviations from this ideal performance we constructed a model which accounts for the dark and background counts and possible mode and phase mis-match between the nulling and signal pulses. The dark and background counts per slot were measured from the direct detection data and additional data taken with the nulling arm on and signal completely blocked. We found that these counts were dominated by leakage from imperfect suppressions in the EOMs and that a probability of dark and background click, $P_{\mathrm{d+bg}}=0.0042 N_p + 0.0129 N_\mathrm{null}$, gave the best fit to the data.   Here $N_\mathrm{null}=\beta^2$ denotes the mean photon number of the nulling pulse. Accounting for this in the model leads to a small correction in the error probability for the DD points at higher $N_p$. This was also included in our model of the CPN receiver and gave similarly small corrections. A more significant effect arises from the fact that the signal and nulling pulses do not perfectly interfere. This modifies the Poisson statistics of the detection events. Most importantly, when attempting to exactly null a pulse, with $\beta=\alpha$, a pulse with mean photon number $\Delta_m N_p$  impinges on the detector rather than vacuum (where $\Delta_m$ is the fractional mode mis-match). We believe  imperfect polarization from our optical elements to be the largest source of such mis-match in our current set-up. We estimate temporal mis-match to be less than 1 $\%$. A phase difference $\theta$ between the signal and nulling pulses in the two arms of the interferometer leads to an equivalent modification of the statistics, where the effective fractional mode mis-match parameter is $\Delta_\theta =2 (1-\mathrm{cos} \, \theta)$. We kept $\Delta_m+\Delta_\theta$ as a free parameter and found that choosing this to be $0.05$ gave fairly good agreement with the observed data points (plotted as the solid red curve in Fig.~\ref{fig:experiment}(a)). We found that for any finite mis-match, the CPN receiver performance begins to degrade at some finite $N_p$.  We expect---with some straight-forward improvements---to be able to achieve $\Delta_m+\Delta_\theta=0.005$, in which case the improvement over DD would be $4.6$~dB at $N_p=2$ photons.

In the current implementation, no active stabilization of the interferometer arms has been implemented, since careful shielding and vibration isolation allow the system to remain stable (to within $3^o$) for the order of seconds, more than the $\sim 0.5$~s required to obtain $832$ frames for each data point. However, the system is designed to accommodate active phase stabilization, in particular, by isolating the signal laser (from a backward-propagating servo laser used for active stabilization) for the detector using multiple bounces on a pair of wavelength-sensitive dichroic mirrors. We note that in a realistic communication system the nulling field will not derive from the same laser as the signal and must be actively stabilized relative to the signal field's phase and mode.

To further improve the error rates, we next implemented and explored the optimized nulling version of the CPN receiver \cite{Guh10} by holding $N_p$ constant and varying $N_\mathrm{null}$. An example with $N_p=0.64$ is plotted in Fig.~\ref{fig:error_rates}(b).  Here we observed an error rate of $0.24$ at the optimal nulling value of $N_\mathrm{null}=1.2$ photons, which is significantly better than the error rate obtained with exact nulling ($N_\mathrm{null} = N_p$) and a $2.1$~dB improvement over DD. This highlights the importance of the optimal nulling generalization of the CPN receiver at lower $N_p$ and is analogous to the improvement the Generalized Kennedy receiver experiences.  We performed nulling photon number sweeps at several $N_p$ and plot the optimal error probabilities as magenta points in Fig.~\ref{fig:error_rates}(a).  We observed a $2.2$~dB improvement over DD at $N_p=1.25$ for the optimal value $N_{\rm null}=1.4$. We note that the data for the nulling photon number sweeps matched best with a somewhat smaller mode-mismatch parameter, $\Delta_m+\Delta_\theta=0.03$ (slightly lower than the best match of $0.05$ for our exact-nulling data set).

\begin{figure}
\centering
\includegraphics[width=\columnwidth]{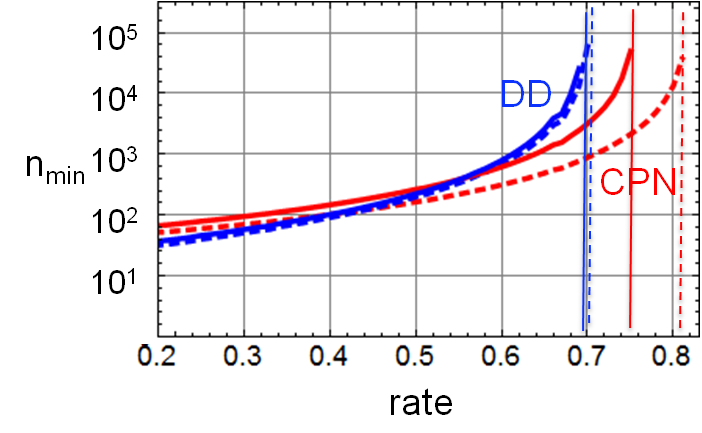}
\caption{Minimum block length $n_{\rm min}$ of a $(n,k,d)$ Reed Solomon outer code which attains a coded block error probability $P_e = 10^{-10}$, as a function of code rate $r=k/n$.  We compare the results of outer coding (over errors and erasures generated by the receiver acting on the inner code) for a $M=4$ PPM inner code with $N_p=1.25$ mean photon number per pulse, both for DD (blue plots), and the CPN receiver (red plots) with optimum nulling $N_\mathrm{null}=1.6$ photons. The dashed curves correspond to the ideal theoretical performance, corresponding to the DD receiver's error and erasure probabilities, $P_e^{(\mathrm{DD})}=0$, $P_{\mathrm{eras}}^{(\mathrm{DD})}=0.289$, and the CPN receiver's error and erasure probabilities, $P_e^{(\mathrm{CPN})}=0.082$, $P_{\mathrm{eras}}^{(\mathrm{CPN})}=0.011$.  The solid curves correspond to our experimentally demonstrated error and erasure rates for the two receiver strategies, $P_e^{(\mathrm{DD})}=0.004, P_{\mathrm{eras}}^{(\mathrm{DD})}=0.287$ and $P_e^{(\mathrm{CPN})}=0.092, P_{\mathrm{eras}}^{(\mathrm{CPN})}=0.052$ respectively.  An erasure for the CPN receiver occurs in the event of a click in all four time slots (the lower branch in all four steps of the decision tree in Fig.~\ref{fig:CPN}(b)), which has zero probability of occurrence for ideal exact nulling with no dark or background counts. Fortunately, this erasure outcome is the most common incorrect decision caused by signal-null  mode-mismatch, which is an easier kind of error to correct by the outer code (than a hard decision error), made by the inner decoder. The vertical lines indicate the Shannon rates for each of the four cases plotted, at which the $n_{\min}$ required shoots to infinity, and above which rate an outer-code-based error correction becomes ineffective (due to the converse of Shannon's noisy channel coding theorem~\cite{Sha48}).}
\label{fig:codederror}
\end{figure}

In practice, very low coded bit error rates on commununication links employing PPM are obtained by employing an outer code---quite often an $(n,k,d)$ Reed Solomon (RS) code---whose code parameters ($n=$ code length, $2^k = $ number of codewords, $d =$ minimum Hamming distance) are chosen optimally based on the order $M$ of the underlying PPM modulation and the mean photon number per PPM pulse, $N_p$. In fact, a suitably chosen RS outer code has been shown to function close to the Shannon capacity of the PPM direct detection channel at low photon flux, with applications to deep space communication~\cite{Moi06}.  To investigate how our observed CPN codeword demodulation error rates can translate into improved coded error performance, we performed  error calculations using a standard formula for minimum distance decoding of RS outer codes \cite{For66}.  Taking our measured optimal nulling CPN and DD performance at $N_p=1.25$, we calculated the minimum code length $n_\mathrm{min}$ required to bring the coded error rate below $10^{-10}$ as a function of RS code rate $r=k/n$, which corresponds to the rate of information communicated per pulse.  The results corresponding to both the ideal theoretical performance (dashed curves) and our observed error rates (solid) are plotted in Fig.~\ref{fig:codederror}.   One subtlety here is that it is optimal for the inner code (PPM) demodulator to record measurement results that are equally likely to correspond to any of the M possible PPM codewords as `erasures' rather than making a hard codeword decision (and most likely making a error) -- as the RS outer code can more effectively correct erasures than errors.   Interestingly nearly all of the DD receiver's incorrect decisions are recorded as erasures (taken to be instances where no clicks occur in any slot) and as a result at low rates $r<0.5$ the DD receiver actually requires slightly smaller code lengths than the ideal CPN receiver, which records more errors and few erasures.  However, it quickly begins to perform worse than the CPN at $r>0.5$ and as the rate approaches the Shannon rate of the DD receiver, its $n_\mathrm{min}$ quickly grows to $> 10 \times$ larger that of the CPN receiver.  There is furthermore a range, between the Shannon rates of the two receivers, where the coded error rate cannot be made small for the DD receiver, while it can be for the CPN receiver.  The solid curves show that a good fraction of this potential advantage can already be realized with the error and erasure rates observed in our experimental demonstration.   This indicates a big potential savings in the coding complexity of a typical deep space laser communication system.   Alternatively, for a given coding complexity, the CPN receiver could achieve the desired coded error rate with smaller power or over larger range.

Recent theoretical results reveal the significance of receivers that use sequential decoding and quantum feedforward---the hallmark of the CPN receiver---for achieving the Holevo capacity of optical communication. Recently, Lloyd et al. and Tan showed that a sequential decoding approach on a random codebook can achieve the ultimate Holevo capacity for any quantum channel~\cite{Llo10, Hui10}. More recently, using Sen's non-commutative union bound for sequential-decoding receivers~\cite{Sen11}, Wilde has shown that a combination of codeword nulling (which we have implemented herein), along with a ``vacuum or not" QND measurement ($\left\{|{\boldsymbol 0}\rangle\langle{\boldsymbol 0}|, {\hat I}_M - |{\boldsymbol 0}\rangle\langle{\boldsymbol 0}|\right\}$) suffice to attain the Holevo capacity~\cite{Wil11a, Wil11b}. The decoder has the receiver ask, through a series of dichotomic von Neumann measurements, whether the output of the channel was the first codeword, the second codeword, etc. This is quite similar in spirit to the CPN receiver, except that it requires quantum non-demolition (QND) ``yes-no" binary projective measurements after each nulling stage, while the CPN receiver annihilates the quantum state of each pulse slot as it progresses through the codeword. Therefore, figuring out how to do the ``vacuum or not" measurement non destructively, and substituting it for the SPD in our CPN setup while using an optimal code (in place of PPM), would suffice to attain the Holevo limit. Furthermore, we recently found the first near-explicit optimal code, one that achieves the Holevo capacity, called the {\em quantum polar code}, whose receiver measurement also works via a sequence of QND binary projective measurements, however with an exponential savings on the total number of measurement stages required compared with the codeword-nulling plus vacuum-or-not-measurement receiver~\cite{Wil11}. 

In summary, we have demonstrated the first optical communications joint detection receiver, the CPN receiver, which performs a joint measurement over PPM codewords, and have shown demodulation probability of error improvement over the traditional direct detection receiver. We have further demonstrated the optimized nulling generalization of this receiver, and with it observed up to $2.2$~dB error rate improvement over DD, the largest improvement over SQL reported to date for an optical receiver. The pulse photon number regime for which we have observed this gain is relevant to deep space optical communication links and we have calculated that in such a system, our observed improvement can translate into big savings in coding complexity. Finally, our results are an important step in the ongoing quest to design and implement joint detection receivers, which can eventually attain the ultimate Holevo limit for reliable communication of classical data over an optical channel.

\begin{acknowledgments}
This work was funded by the DARPA {\it Information in a Photon (InPho)} program under contract\#HR0011-10-C-0159. Discussions with Dr. Masahiro Takeoka, NICT, Japan are gratefully acknowledged.
\end{acknowledgments}

{\bf Competing interests.} The authors declare that they have no competing financial interests.

{\bf Correspondence.} Correspondence and requests for additional materials should be addressed to Dr. Saikat Guha~(email: sguha@bbn.com).


\begin{thebibliography}{99}
\bibitem{Hel76} C. W. Helstrom, {\em Quantum Detection and Estimation Theory}, Academic, New York, (1976). 
\bibitem{Gio04} V. Giovannetti, S. Guha, S. Lloyd, L. Maccone, J. H. Shapiro, H. P. Yuen, ``Classical capacity of the lossy bosonic channel: the exact solution", Phys. Rev. Lett. {\bf 92}, 027902 (2004).
\bibitem{Hol98} A. S. Holevo, ``The capacity of the quantum channel with general signal states", IEEE Transactions on Information Theory, 44(1):269--273, January (1998).
\bibitem{Hau96} P. Hausladen, R. Jozsa, B. Schumacher, M. Westmoreland, W. K. Wootters, ``Classical information capacity of a quantum channel", Phys. Rev. A {\bf 54}, 1869 (1996).
\bibitem{Dol76} S. Dolinar Jr., ``An optimal receiver for the binary coherent state quantum channel", Research Laboratory of Electronics, MIT, Quarterly Progress Report No. 111, 115--120 (1973). 
\bibitem{Coo07} R. L. Cook, P. J. Martin, and J. M. Geremia, ``Optical coherent state discrimination using a closed-loop quantum measurement", Nature (London) {\bf 446}, 774 (2007). 
\bibitem{Guh11} S. Guha, ``Structured Optical Receivers to Attain Superadditive Capacity and the Holevo Limit", Phys. Rev. Lett. 106, 240502 (2011).
\bibitem{Guh11a} S. Guha, Z. Dutton and J. H. Shapiro, ``On quantum limit of optical communications: concatenated codes and joint-detection receivers," IEEE International Symposium on Information Theory (ISIT) 2011, July 31-August 5, 2011, St. Petersburg, Russia
\bibitem{Guh10} S. Guha, J. L. Habif, M. Takeoka, Jour. of Mod. Opt., ``Approaching Helstrom limits to optical pulse-position demodulation using single photon detection and optical feedback", Journal of Modern Optics, {\bf 58}, Nos. 3-4, 257--265, (2011). 
\bibitem{Dol83} S. J. Dolinar, Jr., ``A near-optimum receiver structure for the detection of $M$-ary optical PPM signals", The Telecommunications and Data Acquisition Progress Report 42� 72: December 1982; NASA: Pasadena, CA, (1983).
\bibitem{Llo10} S. Lloyd, V. Giovannetti, L. Macconne, ``Sequential projective measurements for channel decoding", Phys. Rev. Lett. 106, 250501 (2011).
\bibitem{Wil11} M. Wilde and S. Guha, ``Polar codes for classical-quantum channels", {\em submitted to IEEE Trans. Inf. Th.}, arXiv:1109.2591v1 [quant-ph], (2011).


\bibitem{Ken72} R. S. Kennedy, ``A near-optimum receiver for the binary coherent-state quantum channel", Research Laboratory of Electronics, MIT, Quarterly Progress Report No. 110, 219--225 (1972). 
\bibitem{Tak08} M. Takeoka and M. Sasaki, Phys. Rev. A, {\bf 78}, 022320, (2008).
\bibitem{Wit08} C. Wittmann, M. Takeoka, K. N. Cassemiro, M. Sasaki, G. Leuchs, and U. L. Andersen, ``Demonstration of near-optimal discrimination of optical coherent states", Phys. Rev. Lett. {\bf 101}, 210501 (2008). 
\bibitem{Bon93} R. S. Bondurant, Opt. Lett., {\bf 18}, 22, 1896–-1898, (1993).

\bibitem{Sha48} C. E. Shannon, {\em Bell System Technical Journal}~{\bf 27}, (1948).

\bibitem{Sas98} M. Sasaki, K. Kato, M. Izutsu, O. Hirota, ``Quantum channels showing superadditivity in classical capacity", Phys. Rev. A {\bf 58}, 146 (1998).
\bibitem{Fuc97} J. R. Buck, S. J. van Enk, C. A. Fuchs, ``Experimental proposal for achieving superadditive communication capacities with a binary quantum alphabet", Phys. Rev. A {\bf 61}, 032309, (2000).
\bibitem{Tsu11} K. Tsukino, D. Fukuda, G. Fijii, S. Inoue, M. Fujiwara, M. Takeoka, M. Sasaki, ``Quantum receiver beyond the standard quantum limit of coherent optical communication,'' Phys. Rev. Lett. {\bf 106}, 250503 (2011).
\bibitem{Moi06} B. Moision, J. Hamkins, M. Cheng, ``Design of a Coded Modulation for Deep Space Optical Communications", (2006).
\bibitem{For66} G. D.~Forney, ``Concatendated Codes'', Appendix D, Research Monograph No. 37, MIT Press (Cambridge, MA 1966).  
\bibitem{Hui10} S. H. Tan, MIT Ph.D. Thesis, (2011).
\bibitem{Wil11a} M. Wilde, {\em personal communication}, (2011).
\bibitem{Wil11b} M. Wilde, ``From Classical to Quantum Shannon Theory", Exercise 19.3.5, page 454, arXiv:1106.1445v2 [quant-ph], (2011). 
\bibitem{Sen11} P. Sen, ``Achieving the Han-Kobayashi inner bound for the quantum interference channel by sequential decoding", arXiv:1109.0802v1 [quant-ph], (2011).

%\bibitem{Sho00} P. Shor, ``On the Number of Elements Needed in a POVM Attaining the Accessible Information", arXiv:quant-ph/0009077v1, (2000).
%\bibitem{For66} G. D. Forney, {\em Concatenated Codes}, MIT Press, Cambridge, MA, (1966).
%\bibitem{Sef10} S. Sefi and P. Loock, ``How to decompose arbitrary continuous-variable quantum operations", Phys. Rev. Lett. 107, 170501 (2011).


\end{thebibliography}
\end{document}